\documentclass[times,sort&compress]{elsarticle}
\usepackage{hyperref}

\begin{document}

\title{Roles of the spreading scope and effectiveness in spreading dynamics on multiplex networks}

\author[ustc1]{Ming Li\corref{cor1}}
\ead{minglichn@ustc.edu.cn}
\author[hnu]{Run-Ran Liu}
\author[cas]{Dan Peng}
\author[hnu]{Chun-Xiao Jia}
\author[ustc2]{Bing-Hong Wang}

\cortext[cor1]{Corresponding author}

\address[ustc]{School of Engineering Science, University of Science and Technology of China, Hefei, 230026, People's Republic of China}
\address[hnu]{Alibaba Research Center for Complexity Sciences, Hangzhou Normal University, Hangzhou, 310036, People's Republic of China}
\address[cas]{Anhui Province Key Laboratory of Medical Physics and Technology, Center of Medical Physics and Technology, Hefei Institutes of Physical Science, Chinese Academy of Sciences, 350 Shushanhu Road, Hefei, 230031, People's Republic of China}
\address[ustc2]{Department of Modern Physics, University of Science and Technology of China, Hefei, 230026, People's Republic of China}

\begin{abstract}
Comparing with single networks, the multiplex networks bring two main effects on the spreading process among individuals. First, the pathogen or information can be transmitted to more individuals through different layers at one time, which enlarges the spreading scope. Second, through different layers, an individual can also transmit the pathogen or information to the same individuals more than once at one time, which makes the spreading more effective. To understand the different roles of the spreading scope and effectiveness, we propose an epidemic model on multiplex networks with link overlapping, where the spreading effectiveness of each interaction as well as the variety of channels (spreading scope) can be controlled by the number of overlapping links. We find that for Poisson degree distribution, increasing the epidemic scope (the first effect) is more efficient than enhancing epidemic probability (the second effect) to facilitate the spreading process. However, for power-law degree distribution, the effects of the two factors on the spreading dynamics become complicated. Enhancing epidemic probability makes pathogen or rumor easier to outbreak in a finite system. But after that increasing epidemic scopes is still more effective for a wide spreading. Theoretical results along with reasonable explanation for these phenomena are all given in this paper, which indicates that the epidemic scope could play an important role in the spreading dynamics.
\end{abstract}


\maketitle

\section{Introduction}

The spreading dynamic is one of the important research fields in network science\cite{RevModPhys.87.925}, which cannot only model the spreading of epidemic, opinion and rumor in our daily life, but also reflect some universal physical properties, such as phase transition and critical phenomena. The models usually used in these studies are the so-called SIR and SIS models. Here, $S$ stands for susceptible, $I$ for infective, and $R$ for removed or recovered. In the spreading process, if an $S$ is adjacent to an $I$, it will become an $I$ in the next time step with an epidemic probability. At the same time, an $I$ could recover and become an $S$ (SIS model), or acquire immunity (or die) and become an $R$ (SIR model) in the next time step with another probability, which often takes value $1$ to simplify the model. These studies focus on the relationship between the number of $I$ (SIS model) or $R$ (SIR model) and the epidemic probability. Mathematically, these researches aim to get the epidemic threshold, above which there is an outbreak of pathogen, opinion or rumor in the system.

In theory, the degree-based mean-field theory is often used to solve SIS model\cite{PhysRevLett.86.3200}, which is developed from the classic method for that in a well-mixed system\cite{bailey1975mathematical}. For SIR model, the bond percolation is one of the prevalent methods\cite{PhysRevE.66.016128}, which considers the emergence of the giant component connected by occupied links, when each link is occupied with a probability $T$. Here, the giant component is a connected component that contains a constant fraction of the entire network's nodes\cite{randomgraphs}. Obviously, for a probability $T$, if there is no giant component in the network, the pathogen, opinion or rumor cannot spread widely in the network for an SIR model with epidemic probability $T$. Therefore, the critical point $T_c$ of the bond percolation on the network is also the epidemic threshold of the corresponding SIR model\cite{PhysRevE.66.016128}. In simulations, instead of evolving the system step by step as the spreading mechanism required, we can also use the bond percolation to model the spreading process for a lower time complexity. Besides, there are some other theoretical approaches to study spreading dynamics on networks, one can refer to the recent review article\cite{0034-4885-80-3-036603}.

In reality, the spreading process in a group of individuals often contains more than one layer of connections, such as the spreading of opinion or rumor online and offline. To study this complex spreading process, the models on multiplex networks have been proposed\cite{Boccaletti20141}. In these models, nodes can interact with each other through different connections, which are represented by the links in different layers. The previous studies mainly consider the interaction of pathogens or information among individuals through different layers, such as two pathogens with mutual exclusion mechanism\cite{PhysRevE.81.036118,PhysRevE.84.026105}, spreading of an epidemic and information awareness to prevent infection\cite{PhysRevE.90.012808,PhysRevE.90.052817,PhysRevE.91.012822,wangtang}, collaborating epidemic\cite{PhysRevE.90.062803}, immunization strategy\cite{0295-5075-109-2-26001}, and cooperative epidemics\cite{PhysRevE.93.042303}. Although the modeling methods for these critical issues may be different, the bond percolation has been treated as a mainstream approach for theoretical analysis\cite{PhysRevX.6.021002}. The findings of these works indicate that the multi-interaction plays an important role in the spreading dynamics.

\begin{figure}
\begin{center}
\includegraphics[width=0.6\textwidth]{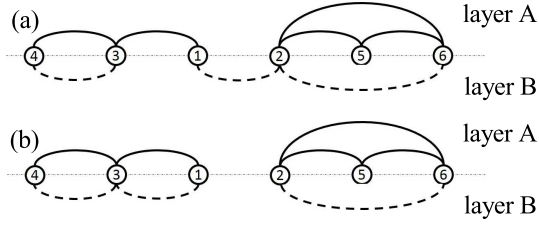}
\caption{A simple graphical representation of the spreading dynamics on multiplex networks with link overlapping. There is only one set of nodes in the system, and the links of the two layers $A$ and $B$ are represented by dash and solid lines, respectively. Assuming that the degrees of the nodes are all fixed. (a) Node $1$ can transmit pathogen to two different nodes (nodes $2$ and $3$) with probabilities $T^A$ (layer $A$) and $T^B$ (layer $B$), respectively. (b) Node $1$ can transmit pathogen to the same node (node $3$) through different layers with probabilities $T^A$ (layer $A$) and $T^B$ (layer $B$), respectively. Overall, the probability is $1-(1-T^A)(1-T^B)=T^A+T^B-T^AT^B$. It is clear that the infection rate of node $3$ in case (b) is larger than that of case (a), however, in case (b) node $2$ will never be infected by node $1$.} \label{f1}
\end{center}
\end{figure}

Comparing with simple networks, the multiplex network brings two main effects on the spreading process among individuals as shown in Fig.\ref{f1}. First, the pathogen or information can be transmitted to more individuals through different layers at one time, which increases the number of individuals that will likely become infected or informed. Second, through different layers, an individual can transmit the pathogen or information to the same individuals more than once at one time, which increases the success rate of spreading. This is to say that the infection probabilities could be heterogeneous, which on average retards the spreading\cite{Qu2017}. To get a better understanding of the two competitive effects, i.e., the effects of the diversity of the spreading scopes and effectiveness, we study a spreading process on multiplex networks with link overlapping in this paper. By adjusting the fraction of overlapping, we can get the differences and combined effects of the epidemic scope and effectiveness in the spreading dynamics on multiplex networks. Note that this model is different with the epidemics on interconnected networks\cite{PhysRevE.85.066109,PhysRevE.86.026106,PhysRevE.88.022801} or coupled networks with node overlapping\cite{10.1371/journal.pone.0092200,6517108}. In those models, the nodes for different layers are also different and the overlapping refers to nodes. That is the main difference between multiplex networks and interconnected networks.

\section{Model}

For convenience, two layers of the multiplex network used in our model are labeled as $A$ and $B$, respectively. We assume that the pathogen spreads from a node to its neighbors according to SIR model. That is, a node can be in one of the three states: susceptible ($S$) for the ones that not yet infected, or infectious ($I$) for the ones have already been infected and can transmit the pathogen to other nodes, or removed ($R$) for the ones that died. The pathogen can spread through the links of any of the two layers with the same consequences, i.e., infected or not. To characterize the different efficiencies of the two layers, we assume that $I$ nodes can transmit the pathogen to their $S$ neighbors in layers $A$ and $B$ with probabilities $T^A$ and $T^B$, respectively. In addition, $I$ nodes become $R$ nodes in the next time step.

We assume that a fraction $\beta$ of links in layer $A$ overlap with the links of layer $B$. For two nodes connected directly in both layers, i.e., connected by an overlapping link, the pathogen can be transmitted between the two nodes twice. In this way, the corresponding epidemic probability can be expressed by an overall probability $1-(1-T^A)(1-T^B)=T^A+T^B-T^AT^B$. As shown in Fig.\ref{f1}, the overlapping corresponds to the effect of enhancing the epidemic effectiveness, and the non-overlapping corresponds to the effect of increasing the spreading scopes. In other words, the overlapping has both positive and negative influences in each step. By adjusting the link overlapping fraction $\beta$ of a multiplex network with given degrees, we can find the features of the two aspects. In addition, note that this is different with the model studied in ref.\cite{PhysRevX.6.021002}, in which all the layers have the same link occupied probability. More difference is that they are focused more on critical phenomena and its relation with that in single network.

\section{Theory}

As pointed above, the epidemic model can be mapped into a bond percolation. The only difference is that there are two different occupied probabilities $T^A$ and $T^B$ for the two layers, respectively. Assuming that the average degrees of the two layers are $z^A$ and $z^B$. Then, the average degree of the overlapping links is $\beta z^A$. Excluding these links, the average degrees of layers $A$ and $B$ are $(1-\beta)z^A$ and $z^B-\beta z^A$, respectively. In what follows, we use script $a$, $b$ and $ab$ to distinguish the parameters for the three types of links, i.e., $a$ for the links in layer $A$ excluding the overlapping links, $b$ for the links in layer $B$ excluding the overlapping links and $ab$ for the overlapping links.

Next, let us solve this bond percolation problem. As mentioned above, it is straightforward that if a node belongs to the giant component, at least one of its links (any types) must be occupied and connect to the giant component. This indicates that the fraction of the nodes in the giant component $\psi$ can be written as
\begin{equation}
\psi = 1-\sum_{k^a,k^b,k^{ab}}p_{k^a,k^b,k^{ab}}\left(1-T^A\varphi^a\right)^{k^a}\left(1-T^B\varphi^b\right)^{k^b} \left[1-\left(T^A+T^B-T^AT^B\right)\varphi^{ab}\right]^{k^{ab}}. \label{ss}
\end{equation}
Here, $p_{k^a,k^b,k^{ab}}$ is the joint distribution of the degrees $k^a$, $k^b$ and $k^{ab}$, and $\varphi^a$ ($\varphi^b$ or $\varphi^{ab}$) is the probability that a node, reached by following a link of type $a$ ($b$ or $ab$), belongs to the giant component. It is easy to know that $(1-T^l\varphi^l)^{k^l}$ is the probability that the node cannot connect to the giant component through links of type $l$ ($a$, $b$ or $ab$). In this way, the sum in eq.(\ref{ss}) means that the node cannot connect to the giant component by any type of links.

If the three degrees $k^a$, $k^b$ and $k^{ab}$ are independent of each other, equation (\ref{ss}) can be expressed in a simple form by the generating functions of these degree distributions, $G_0(x)=\sum_kp_kx^k$,
\begin{equation}
\psi = 1-G_0^{a}\left(1-T^A\varphi^a\right)G_0^{b}\left(1-T^B\varphi^b\right)G_0^{ab}\left[1-\left(T^A+T^B-T^AT^B\right)\varphi^{ab}\right]. \label{s}
\end{equation}
Here, the generating functions $G_0^{l}$ gives the probability that a randomly chosen node cannot connect to the giant component by links of type $l$.

To obtain $\psi$, we must get $\varphi^a$, $\varphi^b$ and $\varphi^{ab}$ firstly. Using the generating function of the excess-degree distribution $G_1(x)=\sum_kp_kkx^{k-1}/\sum_kp_kk$, we can write $\varphi^a$, $\varphi^b$ and $\varphi^{ab}$ in a similar form of eq.(\ref{s}),
\begin{eqnarray}
\varphi^a &=& 1-G_1^{a}\left(1-T^A\varphi^a\right)G_0^{b}\left(1-T^B\varphi^b\right) G_0^{ab}\left[1-\left(T^A+T^B-T^AT^B\right)\varphi^{ab}\right],  \label{ra} \\
\varphi^b &=& 1-G_0^{a}\left(1-T^A\varphi^a\right)G_1^{b}\left(1-T^B\varphi^b\right) G_0^{ab}\left[1-\left(T^A+T^B-T^AT^B\right)\varphi^{ab}\right],  \label{rb} \\
\varphi^{ab} &=& 1-G_0^{a}\left(1-T^A\varphi^a\right)G_0^{b}\left(1-T^B\varphi^b\right) G_1^{ab}\left[1-\left(T^A+T^B-T^AT^B\right)\varphi^{ab}\right].  \label{ro}
\end{eqnarray}
Here, the generating functions $G_1^{l}$ gives the probability that the node, reached by following an $l$ link, cannot connect to the giant component by $l$ links. In this way, the sums of the right hand sides of eqs.(\ref{ra})-(\ref{ro}) mean all the excess links of a node reached by following a corresponding link cannot lead to the giant component. These equations hold only for the case that $k^a$, $k^b$ and $k^{ab}$ are independent of each other, or we must write them in a form similar to eq.(\ref{ss}).

In general, we can solve eqs.(\ref{ra})-(\ref{ro}) to obtain $\varphi^a$, $\varphi^b$ and $\varphi^{ab}$, and then insert them into eq.(\ref{s}) to get the order parameter $\psi$. Below the critical point $T^A_c$ or $T^B_c$, all these will lead to a zero $\psi$, corresponding to that the pathogen or rumor dies out.

\section{Simulation results and discussion}

\subsection{Poisson degree distribution}

As an example, we consider the case that the two layers are both Erd\H{o}s-R\'{e}nyi (ER) networks with link overlapping. Thus, it is easy to know that all the three degree distributions $p_{k^a}$, $p_{k^b}$ and $p_{k^{ab}}$ used in eqs.(\ref{ra})-(\ref{ro}) follow Poisson distribution. In this case, $G_0(x)=G_1(x)=e^{z(x-1)}$, and $G_0^{\prime}(1)=G_1^{\prime}(1)=z$, so $\psi$, $\varphi^a$, $\varphi^b$ and $\varphi^{ab}$ are equivalent. This yields
\begin{equation}
\psi = 1-e^{-\left(T^Az^A+T^Bz^B-\beta T^AT^Bz^A\right)\psi}. \label{ser}
\end{equation}
There are two control parameters $T^A$ and $T^B$ in this equation, so we will check its solutions from the following two cases.

\begin{figure}
\begin{center}
\includegraphics[width=0.8\textwidth]{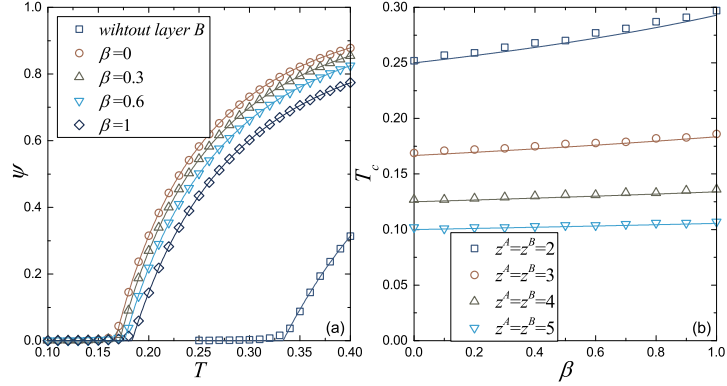} \caption{(color online) (a) The giant component $\psi$ as a function of the occupied probability $T$. The size of the network is $N=10^5$, and the average degrees of the two layers are $z^A=z^B=3$, respectively. The degree distributions $p_{k^a}$, $p_{k^b}$ and $p_{k^{ab}}$ are all Poisson distribution. The corresponding lines are the theoretical results obtained by eq.(\ref{ser}). (b) The critical point of the bond percolation $T_c$ as a function of the overlapping fraction $\beta$. The size of the network is $N=10^6$. The corresponding lines are the theoretical results obtained by eq.(\ref{otc}). } \label{f2}
\end{center}
\end{figure}

\textit{Case $1$ :} $T^A=T^B=T$. For this case, the occupied probabilities $T^A$ and $T^B$ for the two layers are equal to each other. From the simulation results shown in Fig.\ref{f2} (a), we can find that for two given layers, the critical point increases with the overlapping. In other words, the overlapping suppresses the spreading. As the discussion for Fig.\ref{f1}, with the increasing of the overlapping fraction, there are two main effects on the spreading process among individuals, one is enhancing the epidemic effectiveness and the other is reducing the epidemic scope. The first facilitates the spreading locally, and the second suppresses the spreading globally. Together with the results shown in Fig.\ref{f2}, we conclude that adding an additional layer will facilitate the spreading, regardless of the correlation between the original network and the additional layer. However, the best is the one without link overlapping. That is to say that increasing epidemic scope is more efficient than enhancing epidemic effectiveness to facilitate the spreading process.

Approaching the critical point, $\psi \rightarrow 0$, so we can solve eq.(\ref{ser}) in this condition to get the critical point, that is
\begin{equation}
\beta z^A T_c^2 - \left(z^A+z^B\right)T_c + 1=0. \label{tc}
\end{equation}
Obviously, equation (\ref{tc}) gives
\begin{equation}
T_c = \frac{z^A+z^B - \sqrt{\left(z^A+z^B\right)^2-4\beta z^A}}{2\beta z^A}. \label{otc}
\end{equation}
So now, we obtain the critical point of the system. This theoretical result is consistent with the simulation results shown in Fig.\ref{f2} (b). From Fig.\ref{f2} (b), we can also find that the critical point $T_c$ decreases with the increasing of the average degrees $z^A$ and $z^B$. This is quite understandable since more connections will facilitate the spreading process.

\textit{Case $2$ :} $T^A=constant$. Similar with case $1$, we can also solve eq.(\ref{ser}) in the condition $\psi \rightarrow 0$ to obtain the critical point, the only difference is that $T^A$ is a constant. This yields
\begin{equation}
T_c^B = \frac{1-T^Az^A}{z^B-\beta T^Az^A}. \label{tcb}
\end{equation}
From this equation, we can find that for
\begin{equation}
T^A \geq \frac{1}{z^A},
\end{equation}
$T_c^B<0$, which means that there is no percolation transition in this system, and the giant component always exists in this system. In other words, for this epidemic probability $T^A$, the pathogen or rumor can outbreak in the system without the participation of layer $B$, and $T^B$ only affects the outbreak size of the pathogen or rumor. In addition, if $T^B$ is a constant, the situation is similar, we do not repeat here for reason of brevity.

\begin{figure}
\begin{center}
\includegraphics[width=0.8\textwidth]{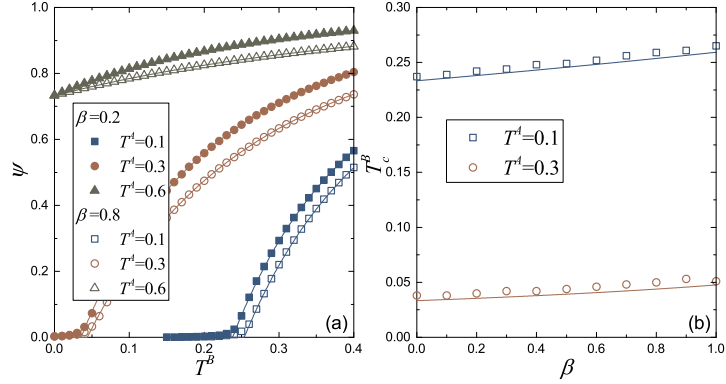} \caption{(color online) (a) The giant component $\psi$ as a function of the occupied probability $T^B$. The size of the network is $N=10^5$. The corresponding lines are the theoretical results obtained by eq.(\ref{ser}). (b) The critical point $T^B_c$ of the bond percolation as a function of the overlapping average fraction $\beta$. The size of the network is $N=10^6$. The corresponding lines are the theoretical results obtained by eq.(\ref{tcb}). The average degrees of the networks used in the simulation are $z^A=z^B=3$. } \label{f3}
\end{center}
\end{figure}

The simulation results for this case are shown in Fig.\ref{f3}, which are in agreement with our analysis well. As our theory predicts, for $T^A=0.6$, there is no percolation transition in the system. From Fig.\ref{f3}, we can also find the similar results with case $1$, that is the link overlapping suppresses the spreading process. For all, it can be summarized as that for Poisson distributions, increasing the epidemic scope is more efficient than enhancing epidemic effectiveness to facilitate the spreading process.

\subsection{Scale-free degree distribution}

For real networks, the degree distribution often takes the form $p_k\sim k^{-\gamma}, 2<\gamma<3$, that is the scale-free network. For this distribution, $G_1^{\prime}(1)$ will be divergency, that leads to $T_c\rightarrow 0$\cite{PhysRevE.66.016128}. For multiplex networks, if one layer takes such a degree distribution, the degree of the overlapping links must obey the same distribution but with a smaller average degree. This results in that we cannot freely choose the degree distribution of the other layers, unless the overlapping links are very few.

\begin{figure}
\begin{center}
\includegraphics[width=0.8\textwidth]{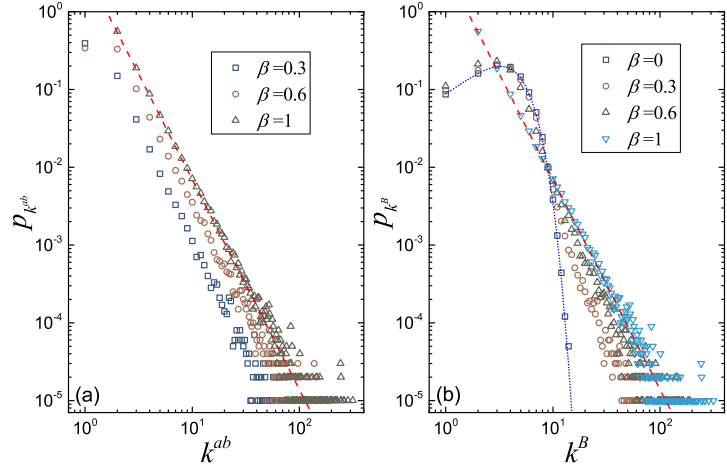} \caption{(color online) The degree distributions of the multiplex network generated by the method presented in the text. (a) The degree distribution of the overlapping links $p_{k^{ab}}$ for different overlapping fractions. (b) The degree distribution of layer $B$, $p_{k^B}$ for different overlapping fractions. In the simulation, layer $A$ takes a scale-free degree distribution $p_{k^A}\sim \left(k^A\right)^{-\gamma}$ with $\gamma=2.7$. The network size is $N=10^5$, and the corresponding average degrees are $z^A=z^B=3.752$. The red dashed line demonstrates the power-law relation with scaling exponent $-2.7$, and the blue dot line follows a Poisson distribution with average $3.752$.} \label{f4}
\end{center}
\end{figure}

To study our model with a scale-free degree distribution, we generate a multiplex network as follow. First, we generate a scale-free network by the configuration model\cite{Newman:2010:NI:1809753} as one layer of the network, i.e., generating links of layer $A$. Then, randomly choosing a fraction $\beta$ of these links to be the overlapping links $ab$. At last, $b$ links can also be generated by the configuration model, here, we use Poisson degree distribution with $z^B=z^A$. Obviously, $p_{k^A}$, $p_{k^a}$ and $p_{k^{ab}}$ follow the same scale-free distribution but different average degrees $z^A$, $(1-\beta)z^A$ and $\beta z^A$ (see Fig.\ref{f4} (a)). However, for layer $B$, the degree distribution $p_{k^B}$ would be a special distribution, which depends on $p_{k^{ab}}$ and $p_{k^b}$. As shown in Fig.\ref{f4} (b), with the increasing of $\beta$, $p_{k^B}$ will turn from a Poisson distribution to a power law.

Before showing the simulation results, let us revisit the theoretical result in the last section. Obviously, equation (\ref{s}) can be rewritten as
\begin{equation}
\psi = 1-G_0^{A}\left[1-T^A\varphi^A-\beta (1-T^A) T^B\varphi^A\right] G_0^b(1-T^B\varphi^b). \label{ssf}
\end{equation}
Here, $G_0^{A}$ gives the probability that the node cannot connect to the giant component by $A$ links, and $G_0^{b}$ is that of $b$ links. Similarly, equations (\ref{ra})-(\ref{ro}) can also be rewritten as
\begin{eqnarray}
\varphi^{A} &=& 1-G_1^{A}\left[1-T^A\varphi^A-\beta (1-T^A) T^B\varphi^A\right] G_0^b(1-T^B\varphi^b), \label{sra} \\
\varphi^{b} &=& 1-G_0^{A}\left[1-T^A\varphi^A-\beta (1-T^A) T^B\varphi^A\right] G_1^b(1-T^B\varphi^b). \label{srb}
\end{eqnarray}
For scale-free degree distributions, we cannot write the generating functions eqs.(\ref{ssf})-(\ref{srb}) into a simple form like that of Poisson degree distribution. However, for a finite network, there must be a cutoff for the series in the generating function, so the numerical value of each generating function can be obtained easily. In this way, we can also get the theoretical results for such networks with given sizes. In addition, since $p_{k^A}$ is a scale-free distribution, a divergency $G_1^{\prime}(1)$ will be involved when we expand eqs.(\ref{sra}) and (\ref{srb}) near the critical point ($\varphi^A\rightarrow 0$, $\varphi^b\rightarrow 0$). This is to say our model will also give a critical point $T_c\rightarrow 0$ for $N\rightarrow \infty$.

\begin{figure}
\begin{center}
\includegraphics[width=0.8\textwidth]{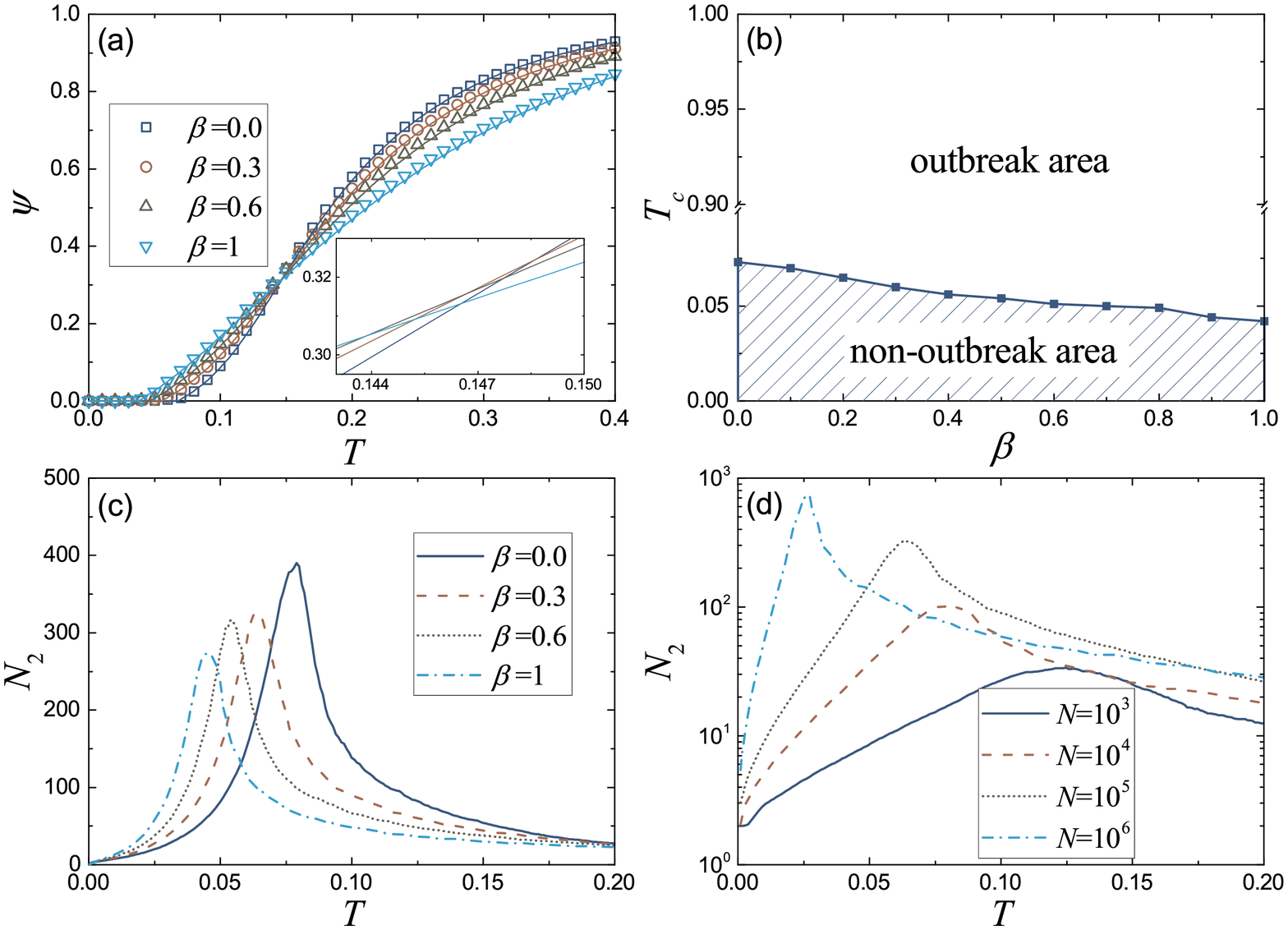} \caption{(color online) The bond percolation on multiplex networks with scale-free degree distribution. The networks are generated as the method presented in the text. The degree distribution of layer $A$ satisfies $p_{k^{A}}\sim \left(k^A\right)^{-\gamma}$ with $\gamma=2.7$. (a) The giant component $\psi$ as a function of the occupied probability $T=T^A=T^B$. The network size is $N=10^5$, and the corresponding average degrees are $z^A=z^B=3.752$. The solid line is obtained by eqs.(\ref{ssf})-(\ref{srb}). The inset figure is the theoretical results for the crossing points of $\psi$ with different $\beta$. (b) The phase diagram of the system, i.e., the pseudo-critical point as a function of the parameter $\beta$. (c) The number of nodes in the second largest component $N_2$ near the critical point for different overlapping fractions. (d) The number of nodes in the second largest component $N_2$ near the critical point for different network sizes. The overlapping fraction is $\beta=0.3$.} \label{f5}
\end{center}
\end{figure}

In Fig.\ref{f5}, we give the simulation results and corresponding theoretical results for the case $T^A=T^B=T$, which agree with each other very well. Different from networks with Poisson degree distributions, a larger overlapping fraction $\beta$ does not always lead to a larger giant component $\psi$. Above some epidemic probabilities, the overlapping suppresses the spreading process as that found in the last subsection. However, below that, the overlapping facilitates the spreading.

As we know, the bond percolation on scale-free networks gives a critical point tending to $0$ when $N\rightarrow \infty$\cite{RevModPhys.87.925}. For networks with finite sizes, the critical point indicated by the maximum point of the second largest component in the network (see Fig.\ref{f5} (c)), called pseudo-critical point, will be very small, but not $0$. Generally speaking, this is because that the hub nodes in these networks can expand the spread scope easily. However, the fundamental premise is that the pathogen or rumor cannot die out in the first several steps, before the the hub nodes is infected. In this way, a local larger epidemic probability is needed to cause pathogen or rumor outbreaks. That is to say that the overlapping facilitates the spreading. i.e., the larger the overlapping fraction $\beta$ is, the smaller the pseudo-critical point is (see Fig.\ref{f5} (b)). In addition, as shown in Fig.\ref{f5} (d), the pseudo-critical point tends to zero for infinite networks. This results that the non-outbreak area in Fig.\ref{f5} (b) will vanish with the increasing of the network size $N$.

For large epidemic probabilities, the pathogen or rumor cannot easily die out. So for wide spreading, the epidemic scope of each node becomes important in a global perspective. As shown in Fig.\ref{f5} (a), when epidemic probability exceeds some values, the overlapping suppresses the spreading process, i.e., for the same epidemic probability, the system with larger overlapping fraction $\beta$ gives a smaller giant component $\psi$. Note that the crossing points are not fixed for different $\beta$ (see the inset figure of Fig.\ref{f5} (a)).

For all the results shown in Fig.\ref{f5}, we summarize that the pathogen or rumor will spread more easily over the multiplex network, if one of the layers takes a scale-free degree distribution, regardless of the degree distribution of the other layer. In a finite system, the link overlapping will make the outbreak of pathogen or rumor easier. This is because that the hubs in such networks have already provided many connection for spreading, which also indicates that the epidemic scope is important for spreading dynamics on multiplex networks.

\section{Conclusion}

In this paper we have studied an epidemic model on multiplex networks with link overlapping, in which a pathogen or rumor can spread among nodes through two types of connections with different epidemic probabilities. Comparing with single networks, in such networks the pathogen or information can be transmitted to more individuals at one time, and an individual can also transmit the pathogen or information to the same individuals twice. The first increases the spreading scope of a node, and the second increases the success rate of spreading between two nodes.

Through simulation and theoretical studies, we find that for Poisson degree distribution, increasing the spreading scope is more efficient than enhancing the epidemic effectiveness to facilitate the spreading process. However, for power-law degree distribution, we find that enhancing the epidemic probability are more effective, which gives a smaller pseudo-critical point. This is because that the hub nodes in such networks have already provided many connections for increasing the spreading scopes. For large-scale spreading, the epidemic just needs to avoid dying out in the first several steps. Once the hub nodes is infected, it becomes hard to prevent the outbreak of the epidemic. This does not indicate that the spreading scope is not important, but the power-law degree distribution provides a particular structure, which optimizes the spreading scope of all nodes as a whole. From another perspective, the overlapping of links also enables the epidemic probabilities between different nodes to be different. So the results also provides helpful insight into understanding the effects of the heterogeneity of the epidemic probabilities.

For these results, we can conclude that the spreading scope may play a more important role in the spreading process than the epidemic effectiveness. If there are enough or proper connections, a low infection pathogen or an incredible rumor can also spread widely in the social network. We think this finding will be helpful for the understanding of the spreading dynamics on real-world multiplex networks.

\section*{Acknowledgments}

The research of M.L. was supported by the National Natural Science Foundation of China under Grant No.61503355. The research of R.-R.L. and C.-X.J. were supported by the National Natural Science Foundation of China under Grant Nos. 61773148 and 61403114 and the Zhejiang Provincial Natural Science Foundation of China under Grant No.LQ14F030009.

\bibliographystyle{elsarticle-num}
\bibliography{ref}

\begin{thebibliography}{10}
\expandafter\ifx\csname url\endcsname\relax
  \def\url#1{\texttt{#1}}\fi
\expandafter\ifx\csname urlprefix\endcsname\relax\def\urlprefix{URL }\fi
\expandafter\ifx\csname href\endcsname\relax
  \def\href#1#2{#2} \def\path#1{#1}\fi

\bibitem{RevModPhys.87.925}
R.~Pastor-Satorras, C.~Castellano, P.~Van~Mieghem, A.~Vespignani, Epidemic
  processes in complex networks, Rev. Mod. Phys. 87 (2015) 925--979.

\bibitem{PhysRevLett.86.3200}
R.~Pastor-Satorras, A.~Vespignani, Epidemic spreading in scale-free networks,
  Phys. Rev. Lett. 86 (2001) 3200--3203.

\bibitem{bailey1975mathematical}
N.~T.~J. Bailey, The mathematical theory of infectious diseases and its
  applications, Charles Griffin \& Company Ltd, London, 1975.

\bibitem{PhysRevE.66.016128}
M.~E.~J. Newman, Spread of epidemic disease on networks, Phys. Rev. E 66 (2002)
  016128.

\bibitem{randomgraphs}
B.~Bollob{\'a}s, Random Graphs, 2nd Edition, Cambridge University Press, 2001.

\bibitem{0034-4885-80-3-036603}
W.~Wang, M.~Tang, H.~E. Stanley, L.~A. Braunstein, Unification of theoretical
  approaches for epidemic spreading on complex networks, Reports on Progress in
  Physics 80~(3) (2017) 036603.

\bibitem{Boccaletti20141}
S.~Boccaletti, G.~Bianconi, R.~Criado, C.~del Genio, J.~G\'omez-Garde{\~n}es,
  M.~Romance, I.~Sendi{\~n}a-Nadal, Z.~Wang, M.~Zanin, The structure and
  dynamics of multilayer networks, Physics Reports 544~(1) (2014) 1--122.

\bibitem{PhysRevE.81.036118}
S.~Funk, V.~A.~A. Jansen, Interacting epidemics on overlay networks, Phys. Rev.
  E 81 (2010) 036118.

\bibitem{PhysRevE.84.026105}
V.~Marceau, P.-A. No\"el, L.~H\'ebert-Dufresne, A.~Allard, L.~J. Dub\'e,
  Modeling the dynamical interaction between epidemics on overlay networks,
  Phys. Rev. E 84 (2011) 026105.

\bibitem{PhysRevE.90.012808}
C.~Granell, S.~G\'omez, A.~Arenas, Competing spreading processes on multiplex
  networks: Awareness and epidemics, Phys. Rev. E 90 (2014) 012808.

\bibitem{PhysRevE.90.052817}
E.~Massaro, F.~Bagnoli, Epidemic spreading and risk perception in multiplex
  networks: A self-organized percolation method, Phys. Rev. E 90 (2014) 052817.

\bibitem{PhysRevE.91.012822}
Q.~Guo, X.~Jiang, Y.~Lei, M.~Li, Y.~Ma, Z.~Zheng, Two-stage effects of
  awareness cascade on epidemic spreading in multiplex networks, Phys. Rev. E
  91 (2015) 012822.

\bibitem{wangtang}
W.~Wang, Q.-H. Liu, S.-M. Cai, M.~Tang, L.~A. Braunstein, H.~E. Stanley,
  Suppressing disease spreading by using information diffusion on multiplex
  networks, Scientific Reports 6 (2016) 29259.

\bibitem{PhysRevE.90.062803}
Y.~Min, J.~Hu, W.~Wang, Y.~Ge, J.~Chang, X.~Jin, Diversity of multilayer
  networks and its impact on collaborating epidemics, Phys. Rev. E 90 (2014)
  062803.

\bibitem{0295-5075-109-2-26001}
C.~Buono, L.~A. Braunstein, Immunization strategy for epidemic spreading on
  multilayer networks, EPL (Europhysics Letters) 109~(2) (2015) 26001.

\bibitem{PhysRevE.93.042303}
N.~Azimi-Tafreshi, Cooperative epidemics on multiplex networks, Phys. Rev. E 93
  (2016) 042303.

\bibitem{PhysRevX.6.021002}
A.~Hackett, D.~Cellai, S.~G\'omez, A.~Arenas, J.~P. Gleeson, Bond percolation
  on multiplex networks, Phys. Rev. X 6 (2016) 021002.

\bibitem{Qu2017}
B.~Qu, H.~Wang, {SIS} epidemic spreading with heterogeneous infection rates,
  IEEE Transactions on Network Science and Engineering 4~(3) (2017) 177--186.

\bibitem{PhysRevE.85.066109}
M.~Dickison, S.~Havlin, H.~E. Stanley, Epidemics on interconnected networks,
  Phys. Rev. E 85 (2012) 066109.

\bibitem{PhysRevE.86.026106}
A.~Saumell-Mendiola, M.~A. Serrano, M.~Bogu\~n\'a, Epidemic spreading on
  interconnected networks, Phys. Rev. E 86 (2012) 026106.

\bibitem{PhysRevE.88.022801}
H.~Wang, Q.~Li, G.~D'Agostino, S.~Havlin, H.~E. Stanley, P.~Van~Mieghem, Effect
  of the interconnected network structure on the epidemic threshold, Phys. Rev.
  E 88 (2013) 022801.

\bibitem{10.1371/journal.pone.0092200}
C.~Buono, L.~G. Alvarez-Zuzek, P.~A. Macri, L.~A. Braunstein, Epidemics in
  partially overlapped multiplex networks, PLoS ONE 9~(3) (2014) 1--5.

\bibitem{6517108}
O.~Yagan, D.~Qian, J.~Zhang, D.~Cochran, Conjoining speeds up information
  diffusion in overlaying social-physical networks, IEEE Journal on Selected
  Areas in Communications 31~(6) (2013) 1038--1048.

\bibitem{Newman:2010:NI:1809753}
M.~E.~J. Newman, Networks: An Introduction, Oxford University Press, New York,
  2010.

\end{thebibliography}

\end{document}